\documentclass[prd,twocolumn,showpacs,amsmath,amssymb]{revtex4}
\usepackage{graphicx}
\begin{document}
\title{Exploring trans-Planckian physics and the curvature effect by primordial power spectrum with WMAP five-year data}
\author{Jie Ren$^1$}
\author{Hong-Guang Zhang$^2$}
\author{Xin-He Meng$^2$}
\affiliation{$^1$Theoretical Physics Division, Chern Institute of
Mathematics, Nankai University, Tianjin 300071, China}
\affiliation{$^2$Department of physics, Nankai University, Tianjin
300071, China}
\date{\today}
\begin{abstract}
The vacuum inflation with a boundary condition specified at a
short-distance scale in a generally primeval non-flat Universe, and
implications of the correspondingly modified primordial power
spectrum via the WMAP data are investigated. We obtain a general
form of the modified primordial power spectrum including the effects
of both possibly new physics scale and the spatial curvature. The
modulation of the primordial power spectrum due to new physics is of
the order $H/\Lambda$, where $H$ is the Hubble parameter during
inflation and $\Lambda$ is the new physics scale that is key to
inflation, while the modulation from the curvature effect is of the
order $K/k^2$, where $K$ is the spatial curvature before inflation
and $k$ is the comoving wave number. We also find that a closed
Universe before inflation is favored by WMAP five-year data.
\end{abstract}
 \maketitle

\section{Introduction}
Great progress achieved in observational cosmology in the past few
years has offered us an opportunity to probe the possible
trans-Planckian physics as one of high energy frontiers in the
spectrum of cosmic microwave background (CMB) radiation
fluctuations. Inflation is increasingly convinced to be a consistent
picture for universe evolution by mainly CMB experiments
\cite{wmap}. The physics at high energy scales is reflected in the
cosmic size magnified by microscopic quantum fluctuations during the
inflation era and the trans-Planckian physics in inflationary
cosmology has been studied and proposed to be probably detectable
\cite{note,eas,map,eas2,bur}. In astrophysics observations the WMAP
data have been analyzed to constrain cosmological models and study
physics during inflation with the primordial power spectrum
\cite{elg03,eas04,mar06}. The imprints of quantum fluctuations
implied by a quantum gravity theory near Planck scale, like
superstring theory is potentially detectable with more precise data
like in the upcoming Planck satellite mission, which will help us to
explore the fundamental laws of nature.

Many aspects of new physics have being studied recently, partly such
as the CPT violation \cite{cpt}, temperature of the inflation
\cite{temp}, modified dispersion relation \cite{dr}, noncommutative
space time \cite{nc}, modified uncertainty principle \cite{up},
relations to dark energy \cite{de} and various other interesting
topics. A simple but well motivated case is that, for an energy
scale $\Lambda$, new physics emerges as a boundary condition in the
momentum space without assuming the specific nature of the
short-distance physics. The effect can be described by a modulation
term to the primordial power spectrum. Some versions of the
modulation of the primordial power spectrum present as a linear term
\cite{note,eas}, while others show that the leading order is
quadratic \cite{qc1,qc2}. Some authors claim that the effects are
linear in $H/\Lambda$ is of crucial importance to have detectable
effects \cite{map} since $H\sim 10^{14}$ GeV during the inflation.
In literature the non-flat pre-inflation Universe has been studied
\cite{curv,curvp} and the modified primordial power spectrum due to
the spatial curvature has been calculated out \cite{curv}, as
another perspective to explore the trans-Planckian physics.

We perform a combined analysis of the modulation to the primordial
power spectrum due to both the new physics scale and the spatial
curvature. Following Danielsson's method \cite{note}, we calculate
the exact solution of the primordial power spectrum, and then
propose an approximate form with explicit physical meanings. The
leading order modulation to the primordial power spectrum by new
physics is a term with order of magnitude $H/\Lambda$, where $H$ is
the Hubble parameter during inflation and $\Lambda$ is the new
physics scale, important for inflation. And the leading order
modulation by the curvature effect is a $K/k^2$ term, where $K$ is
the spatial curvature before inflation and $k$ is the comoving wave
number. Based on this theoretical result, we can add two more
parameters, which describe the new physics scale and the curvature
contributions, into the six-parameter standard model of cosmology
and employ the WMAP five-year data to fit globally the parameters.

This paper is organized as follows: In the next section, we present
the basic calculations of the primordial power spectrum. In Sec III,
we summarize the method and the results obtained by Danielsson and
Easther \textit{et al}. Then we combine the effects of the possibly
new physics scale and the spatial curvature into the primordial
power spectrum. In Sec. IV, we use the newly modified primordial
power spectrum to fit the WMAP data sets. Finally, we present our
conclusions with discussions in the last section.

\section{Basics of the framework}
First considering the Friedmann-Robertson-Walker metric in the flat
space geometry ($K$=0) as the case favored by observational data
\begin{equation}
ds^2=-dt^2+a(t)^2d{\bf x}^2,
\end{equation}
where $a(t)$ is the scale factor. The Hubble parameter is defined by
$H=\dot{a}/a$, where the dot denotes a derivative with respect to
comoving time $t$. In conformal coordinates the metric takes the
form
\begin{equation}
ds^2=a(\eta)^2(-d\eta^2+d{\bf x}^2),
\end{equation}
where $\eta$ is the conformal time. We take $\eta=-\frac{1}{aH}$ as
an approximation if $H$ is near constant during inflation. By
solving Einstein's equation, we obtain the Friedmann equations
\begin{equation}
\frac{3}{a^2}\left(\frac{a'}{a}\right)^2=\frac{8\pi}{M_{\rm
p}^2}\rho,\quad
-\frac{1}{a^2}\left[\frac{2a''}{a}-\left(\frac{a'}{a}\right)^2\right]=\frac{8\pi}{M_{\rm
p}^2}p,
\end{equation}
where $\rho$ and $p$ are respectively the globally isotropic and
homogenous energy density and the pressure for the cosmological
fluid assumed as perfect fluid, and a prime denotes a derivative
with respect to conformal time. The content of the early Universe is
assumed to be described by a scalar field $\phi$ with a potential
$V(\phi)$ from a fundamental theory, and the equation of state (EOS)
can be expressed as $\rho=\frac{1}{2}\dot{\phi}^2+V(\phi)$, with
$p=\frac{1}{2}\dot{\phi}^2-V(\phi)$. The equation of energy
conservation $\dot{\rho}+3H(\rho+p)=0$ for the scalar field is
$\ddot{\phi}+3H\dot{\phi}+V'(\phi)=0$.

The simple inflationary scenario predicts that a single scalar field
generates adiabatic, Gaussian and nearly scale invariant spectra of
scalar and tensor perturbations. We use the rescale field $\mu$ that
is determined by an equation of parametric oscillation
\begin{equation}
\mu_{\rm S,T}''+(k^2-\frac{z_{\rm S,T}''}{z_{\rm S,T}})\mu_{\rm
S,T}=0,\label{eq:mu}
\end{equation}
where $z_{\rm S}=\sqrt{\epsilon}a$ and $z_{\rm T}=a$. Here S and T
denote the scalar and tensor perturbations, respectively, and
$\sqrt{\epsilon}\simeq\dot{\phi}/H$ is the first slow roll parameter
(we take $8\pi/M_{\rm p}\equiv 1$ for simplicity). For the detailed
derivations, see Ref.~\cite{pert}. The solution of Eq.~(\ref{eq:mu})
should satisfy the asymptotic condition
\begin{equation}
\lim_{k/(aH)\to\infty}\mu_{\rm
S,T}(\eta)=\frac{e^{-ik(\eta-\eta_i)}}{\sqrt{2k}}.\label{eq:asym}
\end{equation}
In this paper, we only consider the scalar perturbations and assume
$\epsilon$ is almost constant. The primordial power spectrum for
scalar perturbations is defined by
\begin{equation}
P(k)=\frac{k^3}{2\pi^2}\left|\frac{\mu}{\sqrt{\epsilon}a}\right|^2.
\end{equation}
The scalar spectrum index and the running scalar spectrum index are
defined respectively by
\begin{equation}
n_s-1\equiv\frac{d\ln P(k)}{d\ln k},\quad\alpha\equiv\frac{d^2\ln
P(k)}{d(\ln k)^2}.
\end{equation}

The solution of Friedmann equations combined with the EOS of perfect
fluid is given by
\begin{equation}
a(\eta)\propto\eta^{2/(1+3w)},
\end{equation}
which can be written as $a(\eta)=\eta^{-c}$, if we write the EOS
parameter as $w=-\frac{1}{3}-\frac{2}{3c}$ (which is obtained by a
holographic dark energy model \cite{mli}). Note that both the
power-law inflation $a(t)\propto t^p$ and the exponential inflation
$a(t)\propto e^t$ in terms of comoving time correspond to a
power-law form $a(\eta)\propto\eta^{-c}$ in terms of conformal time.
The case $c=1$, \textit{i.e.}, $w=-1$ the cosmological constant by
definition, is the case that we are most interested in, thus
Eq.~(\ref{eq:mu}) is
\begin{equation}
\mu''+\left[k^2-\frac{c(c+1)}{\eta^2}\right]\mu=0.\label{eq:mu1}
\end{equation}
The exact solution of this equation can be explicitly given out as
\begin{equation}
u(\eta)=C_1\sqrt{\eta}H_{c+1/2}(k\eta)+C_2\sqrt{\eta}H_{c+1/2}^*(k\eta),
\end{equation}
where the integral constants $C_1$ and $C_2$ are determined by the
asymptotic condition Eq.~(\ref{eq:asym}), and $H_l(x)$ is the Hankel
function. Since the asymptotic condition Eq.~(\ref{eq:asym})
describes a spherical wave in momentum space, we choose the
spherical Hankel function $h_l(x)$. The solution should be
$u(\eta)=C(k)\eta\cdot h_c^*(\eta k)$. In the following we can see
that the effect of the new physics is included by a mixture of the
two terms. Thus in the general case, the primordial power spectrum
for scalar perturbations is given by
\begin{equation}
P(k)=\left(\frac{H}{\dot{\phi}}\right)^2\left(\frac{H}{2\pi}\right)^2
\lim_{\eta\to
0}\eta^2|C_1h_c(k\eta)+C_2h_c^*(k\eta)|^2.\label{eq:pps}
\end{equation}
The limit $\eta\to 0$ is because we consider the late times.

\section{New physics scale and the curvature effect}
We would like to discuss the encoded  effects of trans-Planckian
physics in the primordial power spectrum. The main procedure of a
general approach proposed by Danielsson \cite{note} is summarized as
follows. The initial condition is imposed when $p=\Lambda$, where
$\Lambda$ is the energy scale of new physics important for
inflation. We treat $H/\Lambda$ as a free parameter and fit it with
observational data in the next section. Since the comoving momentum
is given by $k=ap=-\frac{\Lambda}{Hk}$, the initial condition is
\begin{equation}
\eta_0=-\frac{\Lambda}{Hk}.\label{eq:np}
\end{equation}
By defining the conjugate momentum
\begin{equation}
\pi_k=\mu_k'-\frac{a'}{a}\mu_k,
\end{equation}
the quantization of the system in terms of time-dependent
oscillators is given by
\begin{eqnarray}
\mu_k(\eta) &=& \frac{1}{\sqrt{2k}}(a_k(\eta)+a_{-k}^\dagger(\eta)),\\
\pi_k(\eta) &=& -i\sqrt{\frac{k}{2}}(a_k(\eta)-a_k^\dagger(\eta)).
\end{eqnarray}
Then we can apply a Bogoliubov transformation to mix the creation
and annihilation operators at a specific conformal time $\eta_0$.
The choice of vacuum
\begin{equation}
a_k(\eta_0)|0,\eta_0\rangle=0
\end{equation}
means that $v_k(\eta_0)=0$ and $\pi_k(\eta_0)=ik\mu_k(\eta_0)$.

We choose the linear combination of the first-order spherical Hankel
function and its conjugate
\begin{equation}
f_k=\frac{A_k}{\sqrt{2k}}e^{-ik\eta}\left(1-\frac{i}{k\eta}\right)
+\frac{B_k}{\sqrt{2k}}e^{ik\eta}\left(1+\frac{i}{k\eta}\right),
\end{equation}
as the solution of Eq.~(\ref{eq:mu1}) in the case $c=1$. The
Wronskian condition
\begin{equation}
\mu_k^*\frac{du_k}{d\tau}-u_k\frac{du_k^*}{d\tau}=-i
\end{equation}
gives
\begin{equation}
|A_k|^2-|B_k|^2=1.
\end{equation}
The choice of vacuum $v_k(\eta_0)=0$ presents
\begin{equation}
B_k=\frac{ie^{-2ik\eta_0}}{2k\eta_0+i}A_k,
\end{equation}
where $A_k\approx 1$ if $\eta_0$ is very large. A more general
calculation for $A_k$ and $B_k$ modified by trans-Planckian physics
can be found in Ref.~\cite{star}. Finally, the primordial power
spectrum is obtained by
\begin{eqnarray}
P(k)&=&\frac{k^3}{2\pi^2a^2}|f_k|^2\nonumber\\
&\sim&\frac{1}{4\pi^2\eta^2a^2}(|A_k|^2+|B_k|^2-A_k^*B_k-A_kB_k^*)\nonumber\\
&=&\left(\frac{H}{2\pi}\right)^2\left[1-\frac{H}{\Lambda}\sin\frac{2\Lambda}{H}
+\frac{H^2}{\Lambda^2}\sin^2\frac{\Lambda}{H}\right].\label{eq:mod1}
\end{eqnarray}
This result is for the gravitational wave, and one needs to take an
extra factor $1/\epsilon$ into account for scalar perturbations.
Easther \textit{et al.} generalize this result to an arbitrary
background \cite{eas}
\begin{equation}
u_k=\frac{1}{2}\sqrt{\frac{\pi}{k}}\sqrt{\frac{y}{1-\epsilon}}
\left[C_+H_\nu\left(\frac{y}{1-\epsilon}\right)+C_-H_\nu^*\left(\frac{y}{1-\epsilon}\right)\right].
\end{equation}
Their result can be written as
\begin{equation}
P(k)=\left[1+\frac{H}{\Lambda}\sin\frac{2\Lambda}{(1-\epsilon)H}
+\frac{H^2}{\Lambda^2}\sin^2\frac{2\Lambda}{(1-\epsilon)H}\right]P_0(k),
\end{equation}
where $P_0(k)$ denotes the primordial power spectrum in Bunch-Davies
vacuum. This is consistent with Danielson's result, since the sigh
before the modulation term is insignificant.

We further generalize the modulation of the primordial power
spectrum to non-flat Universe. The effects of the spatial curvature
before inflation have been studied in Ref.~\cite{curv}. The
calculations for the open Universe and the closed Universe are
slightly different, thus we only show the closed Universe for
example. For the detailed calculations, please see Ref.~\cite{curv}.
Note that a non-zero curvature is assumed before inflation, while
the present curvature is assumed to be zero as universe large
expansion. In a closed Friedmann Universe, the scale factor for a
closed Universe can be solved as
\begin{equation}
a(t)=\frac{\sqrt{K}}{H}\cos Ht,
\end{equation}
where $K$ is the curvature of the space. The scale factor in terms
of the conformal time is
\begin{equation}
a(\eta)=-\frac{\sqrt{K}}{H}\frac{1}{\sin\sqrt{K}\eta},
\end{equation}
where $\frac{\sqrt{K}}{H}<a<\infty$ as
$-\frac{\pi}{2\sqrt{K}}<\eta<0$. Then, the perturbation equation is
\begin{equation}
\mu_k''+\left[k^2-K\left(2\csc^2(\sqrt{K}\eta)+1\right)\right]\mu_k=0.
\end{equation}
The general solution of this equation is given by
\begin{eqnarray}
\mu_k&&=C_1\left[\sqrt{k^2+K}-i\sqrt{K}\cot(\sqrt{K}\eta)\right]e^{-i\sqrt{k^2+K}\eta}\nonumber\\
&&+C_2\left[\sqrt{k^2+K}+i\sqrt{K}\cot(\sqrt{K}\eta)\right]e^{i\sqrt{k^2+K}\eta},
\end{eqnarray}
where
\begin{equation}
C_1=\frac{A_k}{\sqrt{2}(k^2+K)^{3/4}},\quad
C_2=\frac{B_k}{\sqrt{2}(k^2+K)^{3/4}}.
\end{equation}
We use Eq.~(\ref{eq:pps}) to calculate the modulation of the
primordial power spectrum by spatial curvature as
\begin{equation}
P(k)=\frac{1}{(1+K/k^2)^{3/2}}P_0(k).\label{eq:mod2}
\end{equation}

In the following we follow Danielsson's method to encode the new
physics scale that is important for inflation into the primordial
power spectrum for a universe with non-zero curvature. It is easy to
check that the Wronskian condition also gives $|A_k|^2-|B_k|^2=1$.
The choice of vacuum gives
\begin{equation}
B_k=\frac{k-\sqrt{k^2+K}+i\sqrt{K}\cot(\sqrt{K}\eta_0)}{k+\sqrt{k^2+K}+i\sqrt{K}\cot(\sqrt{K}\eta_0)}
e^{-2i\sqrt{k^2+K}\eta_0}A_k.
\end{equation}
We consider the case at late times when $\eta\to 0$ similar to
Eq.~(\ref{eq:mod1}), and obtain the primordial power spectrum as
\begin{eqnarray}
P(k)&=&\left[1+\frac{(k-\sqrt{k^2+K})^2+K\cos(2\sqrt{k^2+K}\eta_0)}{2k\sqrt{k^2+K}}\right.\nonumber\\
&&-\frac{\sqrt{K}\cot(\sqrt{K}\eta_0)\sin(2\sqrt{k^2+K}\eta_0)}{k}\nonumber\\
&&\left.+\frac{K\cot^2(\sqrt{K}\eta_0)\sin^2(\sqrt{k^2+K}\eta_0)}{2k\sqrt{k^2+K}}\right]\nonumber\\
&&\times\frac{1}{(1+K/k^2)^{3/2}}P_0(k),
\end{eqnarray}
which is the final result with the modulation of the primordial
power spectrum in a generally closed Universe. This can be reduced
to Eqs.~(\ref{eq:mod1}) and (\ref{eq:mod2}).

By assuming that corrections from both new physics and the curvature
effect are very small, we can use the linear approximation, in
which, the modified primordial power spectrum is given by
\begin{equation}
P(k)=A_sk^{-2\epsilon}\left[1-\xi
k^{-\epsilon}\sin\left(\frac{2}{\xi
k^{-\epsilon}}\right)-\frac{K}{k^2}\right],\label{eq:mod3}
\end{equation}
where the possibly new physics gives a $\xi=H/\Lambda$ factor
correction to the primordial power spectrum, and the spatial
curvature contributes a $K/k^2$ term, before which the coefficient
can be absorbed in $K$. For the open Universe with negative $K$, we
can also obtain the same form as Eq.~(\ref{eq:mod3}). The physical
meaning is clear: If no new physics, the asymptotic behavior of the
spherical Hankel function is the free spherical wave. The cutoff due
to new physics serves as a boundary condition in momentum space, and
therefore, the choice of the vacuum leads to a mixture of the
positive and negative mode states, \textit{i.e.}, outward- and
inward-going waves classically. The amplitude of the spherical wave
in a non-flat space will be modified and contribute a separate term
by linear approximation. The modulations of the primordial power
spectrum will be magnified in the CMBR fluctuations, which can be
probed by WMAP mission and the future PLANCK satellite observations.

\section{Probing new physics and the curvature effects by WMAP5 data}
Reconstructing the primordial power spectrum \cite{mod,rep,rec} from
CMB experiments is an essential task to understand the early
Universe physics. The commonly assumed form of the primordial power
spectrum is
\begin{equation}
P(k)=A_s\left(\frac{k}{k_0}\right)^{n_s-1},
\end{equation}
where $n_s=1$ corresponds to the Harrison-Zeldovich (scale
invariant) power spectrum. Slow-roll inflationary models predict
that $n_s$ is very close to one. To study more details for the
spectrum, we should also concern on the second-order term. A
standard approach is the logarithmic expansion to $P(k)$ as below
\begin{equation}
\ln P(k)=\ln A_{\rm
s}+(n_s-1)\ln\left(\frac{k}{k_0}\right)+\frac{\alpha}{2}\ln^2\left(\frac{k}{k_0}\right).
\end{equation}

The modification of the power spectrum due to new physics can be
highly model-dependent, nevertheless, we can make as less as
possible assumptions in our model. The modified primordial power
spectrum can be tested with more and more observational CMB data
with increasingly precision. As we know Danielsson's result of the
modification is
\begin{equation}
P(k;\epsilon,\xi)=P_0(k)\left\{1-\xi\left(\frac{k}{k_n}\right)^{-\epsilon}
\sin\left[\frac{2}{\xi}\left(\frac{k}{k_n}\right)^\epsilon\right]\right\}
\end{equation}
where $\xi=4\times 10^{-4}\sqrt{\epsilon}/\gamma$, and
$\gamma=\Lambda/M_{\rm p}$, and the constant $k_n$ corresponds to
the largest scales measurable in the CMB \cite{elg03}. As another
approach, the oscillations in the primordial power spectrum was
studied by setting a step in the inflationary potential \cite{step}.
In the present work, we focus on the curvature effect in the data
fitting.

We propose a general modified primordial power spectrum in form as
\begin{equation}
P(k;\epsilon,\xi,K)=A_s\left(\frac{k}{k_0}\right)^{-2\epsilon}(1-M_\xi-M_K),
\end{equation}
where
\begin{equation}
M_\xi=\xi
\left(\frac{k}{k_0}\right)^{-\epsilon}\sin\left[\frac{2}{\xi}\left(\frac{k}{k_0}\right)^\epsilon\right],\quad
M_K=\frac{K'}{(k/k_0)^2}.
\end{equation}
Here $M_\xi$ denotes the modulation due to new physics, and $M_K$
stands for the modulation from the spatial curvature. We make a data
fitting analysis to the obtained modified primordial power spectrum
by using Monte Carlo techniques as implemented in the publicly
available CosmoMC code \cite{lew} together with the likelihood code
developed by the WMAP team \cite{wmap}. We take $k_0=0.05$
Mpc$^{-1}$.

\begin{figure}[]
\includegraphics{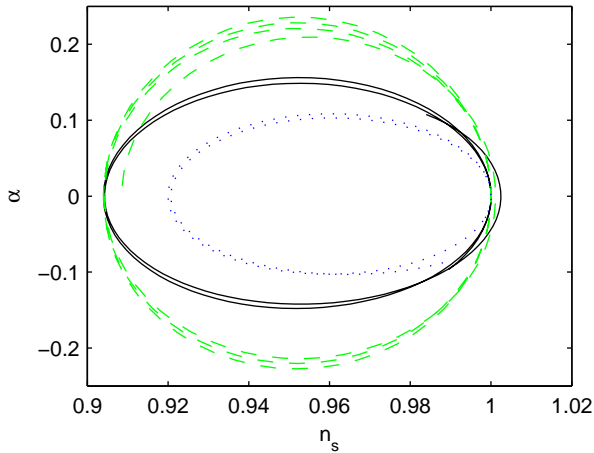}
\caption{\label{fig1} The relation between the scalar spectrum index
and its running. Here we take $K'=10^{-6}$ and the solid, dashed,
and the dotted lines correspond to the parameters ($\epsilon$,
$\xi$) as (0.024, 0.015), (0.024, 0.01), (0.02, 0.015),
respectively.}
\end{figure}
\begin{figure}[]
\includegraphics{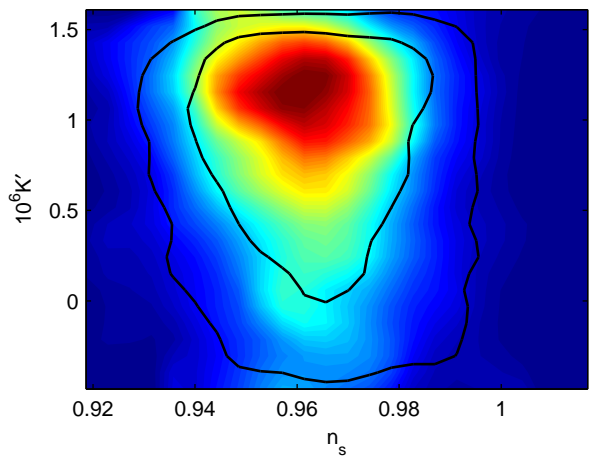}
\caption{\label{fig2} The marginalized contours (68\% and 95\% C.L.)
for the model only with curvature.}
\end{figure}

The parameters in our model are the six parameters in the standard
cosmology model plus two addition parameters that describe the new
physics scale and the spatial curvature, respectively. We have
numerically calculated the scalar spectrum index $n_s$ and its
running $\alpha$ as a function of the wave number $k$,
\textit{i.e.}, $n_s=n_s(k)$ and $\alpha=\alpha(k)$. Since $k$ serves
as a parameter, we cannot obtain the exact values of $n_s$ and
$\alpha$, but we can plot the relation between them. Fig.~\ref{fig1}
plots the relation between $n_s$ and $\alpha$, which is consistent
with the values fitted by the WMAP team. We set $\xi=0$ and fit the
model with data, thus only $K'$ is a new parameter in the model.
Firstly we does restrict the value of $K'$ to be positive or
negative, but after the fitting we find that the closed Universe
with positive $K'$ is favored. Fig.~\ref{fig2} plots the
marginalized contours of the parameters $n_s$ and $K'$. If we take
$K'\sim 10^{-6}$, the energy density of the spatial curvature today
is
\begin{equation}
\Omega_K\sim\frac{K'k_0^2c^2}{a_0^2H_0^2}\sim 10^{-2}.
\end{equation}
In conclusion, the result shows that a model for the primordial
power spectrum with a positive curvature is slightly favored by WMAP
five-year data fittings.

\section{Conclusion and discussion}
We have obtained a general modulation formula to the CMB primordial
power spectrum due to both possibly new physics and the curvature
effect. The leading order of additional terms in the modified
primordial power spectrum is $H/\Lambda$ and $K/k^2$, respectively.
From the classical point of view, the effects of new physics can be
regarded as a reflection result by the boundary of the momentum
space, \textit{i.e.}, a mixture of the outward- and inward-going
waves. The spatial curvature before inflation should not be too
large, otherwise it will largely change the observed power
spectrums, and thus a fine-tuning may be required. Nevertheless, it
is reasonable to include the curvature generally and indeed it
improves the data fitting. We find that a closed pre-inflation
Universe is slightly favored by the data fittings.

Another important issue in cosmology is to understand the nature of
the dark energy dominated current cosmic speed-up expansion. It may
be probable that the inflation and the dark energy can be described
by a unified theory, which exhibits different dynamic and kinematic
behaviors in the early high energy scale universe and the later low
energy scale universe evolutions. Some approaches such as modified
gravity \cite{noj} has been attempted to explore such possibilities.
The observational cosmology has challenged the situation that the
string theory is believed as beyond the experimental science, but
cosmic observations will provide powerful data sets to test the
implications from high energy scale quantum gravity, like string
cosmology \cite{lect}. We can be sure that the upcoming astrophysics
missions, like Planck satellite will open a new era for us to
understand the underlying physics in inflation scenario and dark
energy mystery, with expectations to conceive possibly new physics
and fundamental laws of nature.

\section*{ACKNOWLEDGEMENT}
We acknowledge the use of the Legacy Archive for Microwave
Background Data Analysis (LAMBDA). Support for LAMBDA is provided by
the NASA Office of Space Science. J.R. thanks Prof. Richard Easther
for helpful comments. X.H.M. thanks Prof. David Lyth for helpful
discussions. X.H.M. was supported by NSFC under No.10675062. This
work was supported by Nankai university ISC.

\end{document}